\documentclass[10pt,journal,cspaper,compsoc]{IEEEtran}
\usepackage{graphicx}
\usepackage{subcaption}
\usepackage[disable]{todonotes}
\usepackage{amsmath}
\usepackage{framed}
\usepackage{soul}
\usepackage{url}
\usepackage{multirow}
\usepackage{booktabs}

\begin{document}
%

\title{A Dissection of the Test-Driven Development Process: Does It Really Matter to Test-First or to Test-Last?}



%
%
%

\author{Davide Fucci,~\IEEEmembership{Member,~IEEE,}
        Hakan Erdogmus,~\IEEEmembership{Member,~IEEE}
        Burak Turhan,~\IEEEmembership{Member,~IEEE,}
       Markku Oivo,~\IEEEmembership{Member,~IEEE}
       Natalia Juristo,~\IEEEmembership{Member,~IEEE}
\IEEEcompsocitemizethanks{\IEEEcompsocthanksitem D. Fucci, B. Turhan,  M. Oivo, and N. Juristo are with the Department
of Information Processing Science, University of Oulu, Finland\protect\\
E-mail: {name.surname}@oulu.fi
\IEEEcompsocthanksitem N. Juristo is also with Escuela Tecnica Superior de Ingenieros Informaticos, Universidad Politecnica de Madrid, Spain\protect\\
E-mail: natalia@fi.upm.es
\IEEEcompsocthanksitem H. Erdogmus is with Carnegie Mellon University, Silicon Valley Campus, USA.\protect\\
E-mail: hakan.erdogmus@sv.cmu.edu
}}

\IEEEcompsoctitleabstractindextext{%
\begin{abstract}
\textit{Background}: 
Test-driven development (TDD) is a technique that repeats short coding cycles interleaved with testing. The developer first writes a unit test for the desired functionality, followed by the necessary production code, and refactors the code. Many empirical studies neglect unique process characteristics related to TDD iterative nature.
\textit{Aim}: We formulate four process characteristic: sequencing, granularity, uniformity, and refactoring effort. We investigate how these characteristics impact quality and productivity in TDD and related variations. 
\textit{Method}: We analyzed 82 data points collected from 39 professionals, each capturing the process used while performing a specific development task. We built regression models to assess the impact of process characteristics on quality and productivity. Quality was measured by functional correctness. 
\textit{Result}: Quality and productivity improvements were primarily positively associated with the granularity and uniformity. Sequencing, the order in which test and production code are written, had no important influence. Refactoring effort was negatively associated with both outcomes. We explain the unexpected negative correlation with quality by possible prevalence of mixed refactoring. 
\textit{Conclusion}: The claimed benefits of TDD may not be due to its distinctive test-first dynamic, but rather due to the fact that TDD-like processes encourage fine-grained, steady steps that improve focus and flow.
\end{abstract}

\IEEEkeywords 
Test-driven development, empirical investigation, process dimensions, external quality, productivity.
}

\maketitle
\IEEEdisplaynotcompsoctitleabstractindextext

%

\IEEEpeerreviewmaketitle
     
\section{Introduction}

Test-driven Development (TDD) is a cyclic development technique. Ideally, each cycle coincides with the implementation of a tiny feature. Each cycle finishes once the unit-tests spanning a feature, as well as all the existing regression tests, pass. In TDD, each cycle starts by writing a unit test. This is followed by the implementation necessary to make the test pass and concludes with the refactoring of the code in order to remove any duplication, replace transitional behavior, or improve the design. The cycles can be thought of as iterations of an underlying \textit{micro-process}. 

Advocates of TDD believe that each such cycle, or iteration, should have a short duration and that developers should keep a steady rhythm \cite{jeffries2007guest,beck2003test}. The common recommendation is to keep the  duration of the cycles to five minutes, with a maximum of ten \cite{martin2003agile,jeffries2001extreme}. Thus TDD \textit{cycles} are usually fairly short: we can call them \textit{micro-cycles} or \textit{micro-iterations} to distinguish them from longer notions of cycle or iteration that underlie lifecycle-scale processes, such as sprints in Scrum \cite{schwaber1997scrum}. 

TDD experts also recommend sticking with the prescribed order of the main activities involved: writing a failing test, making the test pass by adding production code, and refactoring \cite{beck2003test}. Note that we distinguish between \textit{test-first} and TDD, with \textit{test-first} referring to just one central aspect of TDD related to the sequencing of the different activities involved in a typical cycle: writing of test code precedes the writing of the production code that the test code exercises. 

Generalizing the above characteristics of a cycle in TDD (length, variation, and order) leads to three basic dimensions: we refer to them as granularity, uniformity, and sequencing. These dimensions do not apply just to TDD but to any cyclic or iterative process. In particular, it applies to several variants that differ from the text-book version in one or more dimensions. Jeffries et al. \cite{jeffries2007guest} see idealized TDD as the endpoint of a step-wise progression from a traditional, monolithic approach towards a modern, iterative one along these dimensions . This progression creates a continuum of processes and gives rise to several viable variants of TDD, such as incremental test-last (ITL) development \cite{erdogmus:encyclopedia}.
\footnote{Although TDD can be as well considered an emerging-design practice, we focus on it as a development technique.}  

Granularity refers to the cycles length, whereas uniformity to their variation, i.e., how constant the duration of the cycles is over time. Granularity and uniformity can be thought of as the tempo and beat in music theory, respectively.

A peculiarity of TDD is in the seemingly counterintuitive sequencing of the activities associated with each cycle: unit tests are written before production code. This particular aspect constitutes the \textit{test-first} component of TDD \cite{beck2003test}. It is captured by the sequencing dimension. This aspect is counterintuitive because the normal software development workflow takes a test-last approach: traditionally, unit testing follows the implementation. It is precisely flipping this component from a test-first to test-last dynamic that gives rise to the ITL variant. Sequencing expresses the preponderance (or lack) of test-first cycles in the development process.

A TDD cycle typically includes an additional activity, refactoring \cite{beck2003test}, in which the structure of the code is improved internally without changing its external behavior detectable by the existing tests \cite{fowler2009refactoring}. Refactoring may involve removal of duplication, other design improvements, or replacement of temporary, stub code. At the end of each refactoring activity, all existing tests should pass, showing that the intended behavior has not been affected by the design changes. However, in practice refactoring is often inter-mixed with other activities, and as such, it sometimes involves adding new production code as well as changing existing tests and behavior. These practical deviations, however, also potentially nullify or reverse the hypothesized benefits \cite{weissgerber2006refactorings, bavota2012does}. 


In summary, the process underlying TDD can be characterized using three main dimensions: granularity, uniformity, and sequencing. The first two dimensions, granularity and uniformity, deal with the iterative, or cyclic, nature of this process. The third dimension, sequencing, deals with the particular activities and their ordering within a cycle.
The test-first nature of TDD is related to this third dimension. 
In addition, a fourth dimension that focuses on refactoring alone captures the prevalence of this activity in the overall process. 

The goal of our paper is to understand, based on the four dimensions, how the elementary constituents of TDD  and related variants affect external software quality and developer productivity. We answer the question: which subset of these dimensions is most salient? This information can liberate developers and organizations who are interested in adopting TDD, and trainers who teach it, from process dogma based on pure intuition, allowing them to focus on aspects that matter most in terms of bottom line. 

This paper is organized as follows: Section \ref{sec:bg} presents the rationale that motivates this study. The settings in which the study took place are described in Section \ref{sec:settings}, and Section \ref{sec:design} presents the study's design. The results are presented in Section \ref{sec:analysis}, the limitations are summarized in Section \ref{sec:ttv}, and related work is reported in Section \ref{sec:background}. The results are discussed in Section \ref{sec:discussion}.
\section{Background and Motivation}
\label{sec:bg}
What makes TDD ``tick'' has been a topic of research interest for some time \cite{jeffries2007guest,Causevic:2011hp,erdogmus2005effectiveness, muller2007effect, Pancur:2011ef, rafique2013effects, wang2004role}. 
In 2014, we set out to investigate this topic rigorously in the context of a larger research project. 
We had collected low level data (Section \ref{sec:metrics}) about the development processes used by our industrial partners.
During the study, in which a test-driven software development process was compared to an iterative test-last (ITL) process, professional software developers tackled programming tasks of different nature (Section \ref{sec:tasks}). 

\begin{figure*}[hbt]
\centering
  \includegraphics[width=.75\textwidth]{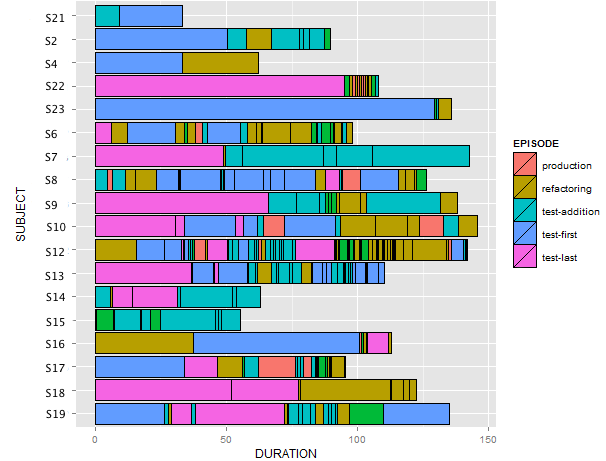}
  \caption{Example visualization of the micro-process underlying the development activities of subjects from a single workshop run. The data were collected from Run 1 subjects in Figure \ref{fig:comparison} applying TDD or ITL to a set of programming tasks. Each row corresponds to a task-subject pair, the subject applying either TDD or ITL to the task (task-technique assignments are not specified in the figure).}
  \label{fig:bar_chart}
\end{figure*}

We visualized the micro-process underlying the data collected from our industrial study on a participant-by-participant basis, as shown in Figure \ref{fig:bar_chart}. We were looking for a metric that would help summarize the process data for each participant in terms of the four dimensions. First we captured each dimension by a quantitative measure easily computable from the data. Then we attempted to combine them into a single \textit{TDD compliance} metric with an eye to regress this metric against the main outcome measures of external quality and developer productivity to assess its potential influence (external quality was equated with functional correctness and developer productivity with speed of production). To do this, three experts independently evaluated each participant's data and gave each resulting observation a TDD compliance score. We then tried to retrofit a formula that would mirror the experts' overlapping assessments. However, not all of the dimensions appeared to have an equally significant contribution. 
Additionally, some dimensions appeared to have a close association with each other. This revelation led us to decide to investigate the four dimensions individually without attempting to aggregate them into an ultimate compliance metric. 

To accomplish our goal, we decided to use data from a set of ongoing quasi-experiments (i.e., studies that lack a control experimental group, and where randomization could not be applied) comparing TDD with ITL. The underlying data of this enclosing context had sufficient variation to allow us to investigate the dimensions of interest using a continuum of micro-processes rather than using the usual dichotomy of clearly delineated treatment and control groups. Thus, this paper is the result of an overlay observational study on top of an experimental context. We call it an \textit{observational study} \cite{mann2003observational}  because by pooling data from all experimental groups, we effectively abstract away from them.
In fact, we have no way of directly controlling the dimensions that serve as factors. The dimensions emerge due to interactions among several factors. A subject's assignment to a particular experimental group with a prescribed process is only one of these factors. Others possibly include  skill, ability, and motivation. Some factors are unknown. The dimensions are not binary in that instead of being present or absent, they are present to varying extents independent of an assigned treatment.  This implies that the dimensions lend themselves best to a regression-type analysis rather than hypothesis testing. 

\begin{figure*}[hbt]
\centering
  \includegraphics[width=0.8\textwidth]{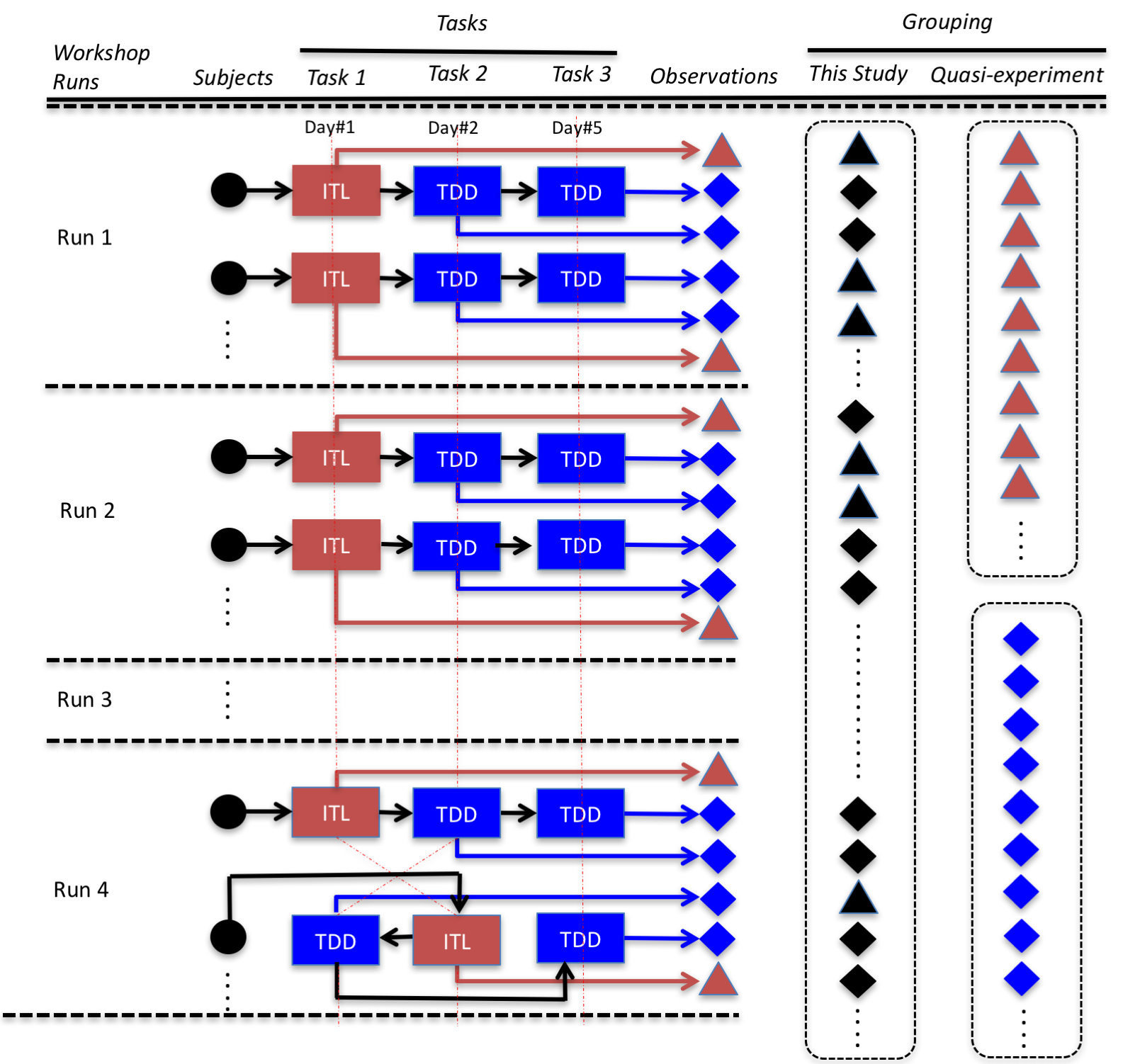}
  \caption{The observations gathered from participants were divided according to the particular technique used by distinct experimental groups formed for the original quasi-experiment (not reported in this paper). In the quasi-experiment, the data were collected from iterative test last (ITL) and test-driven development (TDD) sessions during four runs. Note that Run 4 had a different structure than the other three in terms of the order and development method of the tasks used. For the purpose of the study reported in this paper, however, all the observations from all runs and groups were pooled together, as illustrated under the column labeled "This Study".}
  \label{fig:comparison}
\end{figure*}

Figure \ref{fig:comparison} illustrates how the data gathered from subjects in a repeated-treatment design are used in this study versus in the larger context of the quasi-experiments. The observations are shared between the studies. The outcome variables in both studies are external quality and developer productivity. The sole factor in the experimental context is the treatment group: ITL or TDD. In the current study, all the observations are pooled, and four measures are extracted from each observation to give rise to four factors, each corresponding to one process dimension. 

Ordinarily, observational studies reveal associations: their power to reveal causality is limited. However in our case, there is a difference. The study is overlaid on top of an experimental context: the factors naturally occur in the presence of a deliberate attempt to apply a prescribed process. Similarly the outcome measures, the dependent variables, are observable as a result of the experimental context. Variable omission, or unaccounted confounding factors, is still a concern, but not more than they normally are in an experimental context.
\section{Setting}
\label{sec:settings}
The context of this study is a larger research effort investigating software engineering practices, including test-driven development, in industry \cite{Misirli:2014gk}. This research initiative involves several industrial partners from the Nordic countries. Two companies, referred to as Company A and Company B in the paper, participated in this study. Company A is a large publicly-traded provider of consumer and business security solutions. It is based in Finland, with development centres in Europe, Asia, and the Americas. Company B is an SME based in Estonia that offers on-line entertaining platforms to businesses worldwide. It has development centres in Europe, Asia, and the Middle-East.

This study analyzes aggregate data collected from four runs of a workshop about unit testing and TDD that were conducted at these two companies. In the unit testing portion of the workshop, we taught the participants the principles of unit testing and how to apply unit testing using an iterative process, but based on a conventional test-last dynamic. In the TDD portion of the workshop, we introduced how unit testing is used to drive the development based on a TDD-style test-first dynamic. The workshop runs were held at the participating companies' sites. Three of runs were held at three different sites of Company A (two in Europe and one in Asia), and one run was held in Company B at a European site. Each run lasted five days (eight hours per day, including breaks) in which interactive lectures were interleaved with hands-on practice sessions.
Some of the practice sessions focused on the use of iterative unit testing with a test-last dynamic, which we refer to as \textit{incremental test-last (ITL)}. The other sessions focused on TDD. 
Each participant took part in a single run of the workshop and implemented several tasks, designed for the purpose of the workshop using the techniques that the participant was taught.
We used five tasks to support the teaching in a Randori fashion; the remaining three tasks (see Section \ref{sec:tasks}) were implemented in solo mode, and represent the experimental objects of this study. We also encouraged the participants to try to apply the techniques that they learned during the workshop in their regular work, although we did not collect any data from these informal practice attempts.
The workshop runs were part of a bigger, ongoing multi-object  quasi-experiment on TDD with a repeated-treatment design.

\section{Study Design}
\label{sec:design}
In this section, we state the research questions that are answered in the study. We also elaborate on the choice of the research method, the constructs of interest, the metrics used to measure them, the experimental tasks used, and the data analysis techniques.
\subsection{Research Questions}
We focus on the following research questions with regard to two outcomes--- external software quality and developer productivity---and four factors corresponding to the four process dimensions---sequencing, uniformity, granularity, and refactoring effort. 
\begin{itemize}
	\item \textbf{RQ-QLTY:} Which subset of the factors best  explain the variability in external quality?   
\end{itemize}

\begin{itemize}
	\item \textbf{RQ-PROD:} Which subset of the factors best explain the variability in developer productivity?  
\end{itemize}

The notion of external quality in RQ-QLTY is based on functional correctness, specifically average percentage correctness. We do not consider other notions of external quality, such as usefulness and usability. The notion of productivity in RQ-PROD is based on speed of production, or amount of functionality delivered per unit effort. 

We also consider how the factors interact with each other (for plausible interactions) and whether these interactions matter in predicting the levels of the outcomes. 

\subsection{Method}
We collected fine-grained process data from the four workshop runs to quantify the outcome (dependent) variables and factors (independent variables). For each outcome variable, we use regression analysis to explore its relationship to the four factors and possible interactions among the factors. We also aim to understand the form of any underlying relationships. To have enough variability among the process dimensions represented by the factors, we used pooled data from the four runs, which mixed the application of TDD and ITL.

In our specific case, a controlled experiment
%
was infeasible since it is not possible to control the factors. In this regard, this study is neither a controlled experiment nor a quasi-experiment. No experimental groups (control and treatment groups) can be created by manipulating the factors.  \textit{A priori} random assignment of subjects to factor groups is not possible without such manipulation. 


\subsection{Outcome (Dependent) Variables}\label{sec:dv}
The outcomes under study were external quality, represented by a defect-based (or dually, correctness-based) metric ($QLTY$), and productivity, represented by an effort-based metric ($PROD$). To be able to measure levels of these metrics, we divided the programming tasks used in the workshop runs into many smaller sub-tasks. Each sub-task resembled a small user story. A user story is a description of a feature to be implemented  from the perspective of the end user, and we adopted this perspective when defining the sub-tasks. Because the sub-tasks were very granular, they captured the tasks' functional requirements with high precision. The high level of granularity also made it possible to assess task completion, quality, and correctness on a numerical scale rather than a binary scale.  
The user stories were frozen and known to the subjects upfront: they did not change during the implementation of the tasks by the subjects.
An acceptance test suite---a set of reference tests developed by the researchers---was associated with each user story, with disjoint subsets of acceptance tests in each suite covering specific user stories (see Section \ref{sec:tasks} for details). The acceptance tests were then used to calculate the two outcome metrics, as explained below.
Both metrics, with minor variations, have previously been used in TDD research \cite{erdogmus2005effectiveness,fucci2013role,fucci2014impact,madeyski2006impact}.
\subsubsection{External Quality}
The metric for external quality $QLTY$ measures how well the system matches the functional requirements. Each user story is expressed as a subset of acceptance tests. The formula to calculate $QLTY$ is given in Equation \ref{eq:qlty},
\begin{equation}\label{eq:qlty}
QLTY = \frac{\sum\limits_{i=1}^{\#TUS} QLTY_i}{\#TUS} \times 100,
\end{equation}
where $QLTY_i$ is the quality of the i-th \textit{tackled} user-story, and is defined as in Equation \ref{eq:qltyi}.
\begin{equation}\label{eq:qltyi}
QLTY_i = \frac{\#ASSERT_i(PASS)}{\#ASSERT_i(ALL)}.
\end{equation}
In turn, the number of tackled user stories ($\#TUS$) is defined, similarly to Erdogmus et al. \cite{erdogmus2005effectiveness}, as in Equation \ref{eq:dus}.
\begin{equation}\label{eq:dus}
\#TUS= \sum_{i=0}^{n} 
\begin{cases}
    1 & \#ASSERT_{i}(PASS) > 0\\
    0              & \text{otherwise}
\end{cases}
\end{equation}
where $n$ is the number of user-stories composing the task. $\#ASSERT_i(PASS)$ is the number of passing JUnit assert statements\footnote{http://junit.org/apidocs/org/junit/Assert.html} in the acceptance test suite associated with the $i^{th}$ user-story. Accordingly, a user-story is considered as \textit{tackled} if at least one of its JUnit assert statements passes.
In particular, an empty implementation will not pass any assert statements; conversely, a hard-coded or stubbed implementation may pass one or more assert statements. In the latter case, a user story is still considered as tackled since there is an indication that some work on it has been done.
 We use an assert statement as our smallest unit of compliance with the requirements. This allows us to be able to better highlight quality differences between observations, relative to the enclosing acceptance test case. 
The $QLTY$ measure deliberately excludes unattempted tasks and tasks with zero success; therefore, 
it represents a \textit{local} measure of external quality: it is calculated over the subset of user stories that the subject attempted. $QLTY$ is a ratio measure in the range [0, 100].

For example, suppose that a subject tackles three user stories; i.e., $\#TUS$ = 3. This means there are three user stories for which at least one assert statement passes in the associated acceptance test suite. Say, the acceptance tests of the first tackled user story contains 15 assertions, out of which five are passing; the seconds contains ten assertion, and two are passing; the third also contains ten assertions, and eight are passing. Then the quality value of the first tackled user story is 0.33; the second user story has a quality value of 0.20, and the third 0.80. Therefore, the $QLTY$ measure for the subject is 44\%, the average of the three attempted sub-tasks ($100\times(0.33 + 0.20 + 0.80)/3$).

\subsubsection{Productivity}
The metric for productivity ($PROD$) follows the software engineering definition of \textit{``work accomplished,
with the required quality, in the specified time''} \cite{6263205}.
The metric is calculated as in Equation \ref{eq:prod2} and is defined as a ratio measure in the range is [0, 100]. 
\begin{equation}\label{eq:prod2}
	PROD = \frac{OUTPUT}{TIME}
\end{equation}
$OUTPUT$ (Equation \ref{eq:prod}) is the percentage of assert statements passing for a task when the subject's solution is ran against the whole acceptance test suite for that task. It conveys the proportion of the system requirements that have been correctly implemented. Unlike $QLTY$, this metric is independent of how many of the task's user stories the subject tackled. It does not apply the "at-least-one-assertion" filter to sub-tasks to guess which sub-tasks were attempted and select only those tasks.   
\begin{equation}\label{eq:prod}
OUTPUT = \frac{\#ASSERT(PASS)}{\#ASSERT(ALL)} \times 100,
\end{equation}

$TIME$ (Equation \ref{eq:time}) is an estimate of the number of minutes a subject worked on a task. It was based on log data collected from the subject's IDE. 
\begin{equation}\label{eq:time}
	TIME = \frac{t_{close} - t_{open}}{6000}
\end{equation}
where $t$ denotes a timestamp in millisecond.
Thus $TIME$ is simply the difference, in minutes, between the timestamp of closing the IDE and the timestamp of opening the IDE.
If the subject opended and closed the IDE several times during a run, $TIME$ was calculated as the sum of the all open-close intervals. The maximum time allocated to a task was $\sim$300 minutes, and one minute was the smallest time unit considered.
For example, suppose a subjects implements a task with a total of 50 assert statements in its acceptance test suite. After running the acceptance test suite against the subject's solution, 45 assert statements are passing. Then $OUTPUT = 100 \times 45/50 = 90\%$. Suppose the subject delivered the solution in three hours (i.e., $TIME$ = 180 minutes). The subject's $PROD$ is therefore 0.50, denoting an assertion passing rate of .50\% per minute.  

\subsection{Factors (Independent Variables)}\label{sec:metrics}
\begin{table*}[!ht]
\centering
\caption{Summary of study factors corresponding to the four process dimensions.}
\begin{tabular}{lp{12cm}l}
\hline
Dimension & Interpretation  & Range \\ \hline                                                                                                                                                                                                                                                                  
GRA  & A fine-grained development process is characterized by a cycle duration typically between 5 and 10 minutes. A small value indicates a granular process. A large value indicates a coarse process. & [0, 300]  \\ \hline                                                                  
UNI   & A uniform development process is characterized by cycles having approximately the same duration. A value close to zero indicates a uniform process. A large value indicates a heterogeneous, or unsteady, process. & [0, 149]   \\ \hline                                                             
SEQ  & Indicates the prevalence of test-first sequencing during the development process.  A value close to 100 indicates the use of a predominantly test-first dynamic. A value close to zero indicates a persistent violation of the test-first dynamic. & [0, 100] \\ \hline
REF  & Indicates the prevalence of the refactoring activity in the development process. A value close to zero indicates nearly no detectable refactoring activity (negligible refactoring effort). A value close to 100 indicates a process dominated by refactoring activity (high refactoring effort). & [0, 100]  \\ \hline
\end{tabular}
\label{tbl:dimensions}
\end{table*}
In order to quantify the levels of the four factors---granularity, uniformity, sequencing, and refactoring effort---we decompose the development process applied by a subject into a sequence of small cycles. A cycle is delimited by the successful execution of a regression test suite (the \textit{green bar} in JUnit). Each cycle in turn is composed of a sequence of more elementary actions. The actions form a pattern which help identify a cycle's underlying type of activity---for example, refactoring, test-first production, test-last production, or test addition. This type is inferred automatically by a tool \cite{Becker:2014hz} installed in the IDE and using the heuristics devised in Kou et al. \cite{Kou:2010kv}. 
A cycle also has a duration calculated as the difference between the timestamps of the first and last actions in the cycle.

\begin{figure}[ht]
\centering
	\includegraphics[width=.5\textwidth]{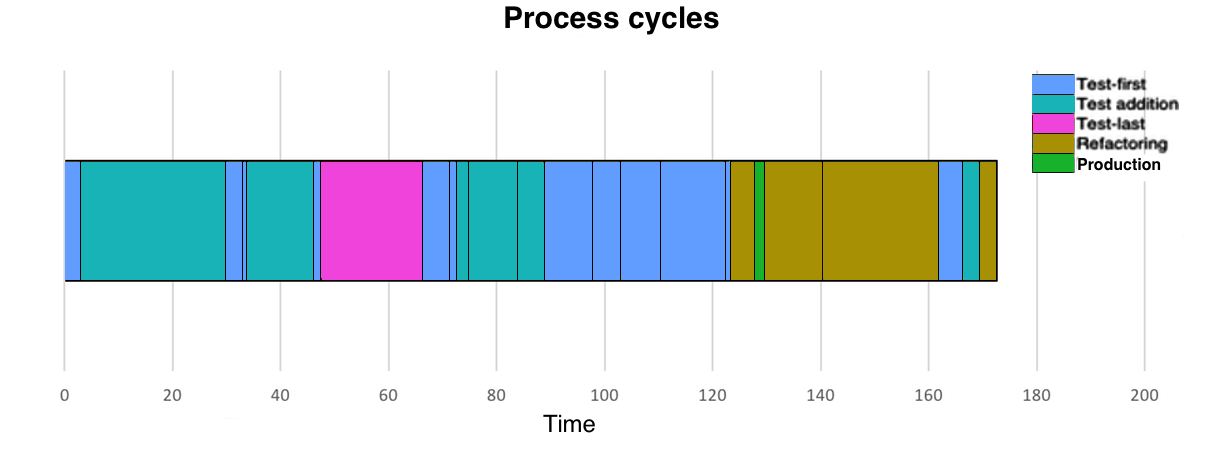}
	\caption{Example visualization of a sequence of development cycles for a single subject working a particular task, equivalent to a single row of Figure \ref{fig:bar_chart}. The width of each section represents the cycle's duration and the color represents the type of the cycle.}
	\label{fig:pulse}
\end{figure}

An example is shown in Figure \ref{fig:pulse}. In the example, time flows from left to right, measured in minutes. Each colored vertical section represents a cycle. The color encodes the type of a cycle. The width of the section represents the cycle's duration. Figure \ref{fig:pulse} captures data collected from a single subject during one of the runs presented in Figure \ref{fig:comparison}. After a first test-first cycle (test-first production, light blue), the subject wrote tests (test addition, aqua) without adding new functionality. New production code was implemented using a test-last approach (test-last production, violet), and after that the subject switched to adding new production code using a test-first dynamic (test-first production again, light blue). The majority of the final cycles were dedicated to refactoring (orange). 

Having characterized the components of the development process in terms of cycle types and durations, we can now relate these components to the four process dimensions to determine their levels for a subject-task pair. Granularity ($GRA$) is measured by the median of cycle duration. Uniformity ($UNI$) is measured by the median absolute deviation (MAD) of cycle duration. We selected median and MAD, respectively, over other typical value and dispersion measures because they are more robust with respect to outliers. Sequencing ($SEQ$) is measured by the fraction of \textit{test-first} cycles, that is, test-first production cycles, which start with a test addition activity, followed by the addition of production code in the middle, and end with successful passing of all tests. Finally, refactoring effort ($REF$) is measured by the fraction of \textit{refactoring} cycles in the development process. In refactoring cycles, normally production or test-code is modified, but no new test or production code is added. Table \ref{tbl:dimensions} summarizes these four process dimensions representing the factors, or independent variables, whereas Table \ref{tbl:heuristics} summarizes the heuristics used by the IDE plug-in to identify the cycle type according to the underlying sequence of actions. In the following section, we explain the computation of the level of each process dimension and the implications of the chosen metrics. 

\begin{table*}[!ht]
\caption{Heuristics used by the tool to infer the type of development cycle (from \cite{Kou:2010kv}).}
\begin{tabular}{ll}
\hline
Type                           & Definition                                                                                                                                               \\ \hline
\multirow{4}{*}{Test-first}    & Test creation $\rightarrow$ Test compilation error $\rightarrow$ Code editing $\rightarrow$Test failure $\rightarrow$ Code editing $\rightarrow$ Test pass \\
                               & Test creation $\rightarrow$ Test compilation error $\rightarrow$ Code editing $\rightarrow$ Test pass                                                        \\
                               & Test creation $\rightarrow$ Code editing $\rightarrow$ Test failure $\rightarrow$ Code editing $\rightarrow$ Test pass                                      \\
                               & Test creation $\rightarrow$ Code editing $\rightarrow$ Test pass                                                                                              \\ \hline
\multirow{3}{*}{Refactoring}   & Test editing (file size changes $\pm$ 100 bytes) $\rightarrow$ Test pass                                                                                                                           \\
                               & Code editing (number of methods, or statements decrease) $\rightarrow$ Test pass                                                                                                                           \\
                               & Test editing AND Code editing $\rightarrow$ Test pass                                                                                                          \\ \hline
\multirow{2}{*}{Test addition} & Test creation $\rightarrow$ Test pass                                                                                                                          \\
                               & Test creation $\rightarrow$ Test failure $\rightarrow$ Test editing $\rightarrow$ Test pass                                                                  \\ \hline

\multirow{2}{*}{Production}& Code editing (number of methods unchanged, statements increase) $\rightarrow$ Test pass \\
								& Code editing (number of methods increase, statements increase) $\rightarrow$ Test pass \\
								& Code editing (size increases) $\rightarrow$ Test pass \\ \hline
\multirow{2}{*}{Test-last}     & Code editing $\rightarrow$ Test creation $\rightarrow$ Test editing $\rightarrow$ Test pass                                                                  \\
                               & Code editing $\rightarrow$ Test creation $\rightarrow$ Test editing $\rightarrow$ Test failure $\rightarrow$ Code editing $\rightarrow$ Test pass          \\ \hline
Unknown                        & None of the above $\rightarrow$ Test pass                                                                                                                      \\ \hline
\end{tabular}
\label{tbl:heuristics}
\end{table*}
\subsubsection{Granularity (GRA)} 
Granularity is measured as the median cycle duration in minutes.
This measure captures the extent of the cyclic nature of the development process used. 
The lower the value, the finer-grained in general the process is. 
A subject with a low $GRA$ value tends to use short cycles, and completes more cycles within the same time than a subject with a higher $GRA$ value. 
We use the median value rather than the average to reduce the sensitivity of the measure to outliers. 
$GRA$ is a ratio measure in the range $[0, 300]$ because the maximum time allocated for a task was 300 minutes.  


\subsubsection{Uniformity (UNI)}
Uniformity represents the dispersion of cycle duration, as measured by mean absolute deviation (MAD).

Uniformity indicates the extent to which a subject was able to keep the \textit{rythm} of the development cycles. 
The lower the value, the more uniform the cycles were. A value of zero indicates that all the cycles had the same duration. UNI (Equation \ref{eq:uni}) is a ratio measure in the range [0, 149], and  calculated as follows:

\begin{equation}\label{eq:uni}
UNI = median(\mid \mathrm{duration}(i) - GRA \mid), i \in [0, n]
\end{equation}
where $n$ is the total number of cycles. 

We use MAD as opposed to standard deviation because $GRA$ is defined as a median. Again this helps reduce the influence of outliers, atypical cycles that are too short or too long. 


\subsubsection{Sequencing (SEQ)}
Sequencing characterizes the underlying development process according to the extent to which a developer adheres to a test-first dynamic. It is measured as the percentage of cycles having type \textit{test-first}.
Note that $SEQ$ does not capture the full nature of TDD, just one central aspect that has to do with its dynamic.\\

$SEQ$ (Equation \ref{eq:seq}) is a ratio measure in the range [0, 100] and it is calculated as follows:
\begin{equation}\label{eq:seq}
SEQ = \frac{\sum_{i=0}^{n} 
\begin{cases}
    1 & \mathrm{type}(i) = \textit{test-first}\\
    0              & \text{otherwise}
\end{cases}
}{n} \times 100,
\end{equation}

where $n$ is the total number of cycles.
\subsubsection{Refactoring Effort (REF)}
Refactoring effort (refactoring in short) is measured by the percentage of cycles likely to be associated with refactoring activity. A cycle recognized by the IDE tool to be an instance of successful refactoring activity has type $refactoring$.

$REF$ (Equation \ref{eq:ref}) is a ratio measure in the range [0, 100], and it is calculated as follows:

\begin{equation}\label{eq:ref}
REF = \frac{\sum_{i=0}^{n} 
\begin{cases}
    1 & \mathrm{type}(i) = \textit{refactoring}\\
    0              & \text{otherwise}
\end{cases}
}{n} \times 100,
\end{equation}
where $n$ is the total number of cycles.

\subsection{Study Tasks}\label{sec:tasks}
The participants implemented three tasks during each workshop run. 
The first two tasks were greenfield: they involved implementing a solution from scratch.
The third task was brownfield: it involved extending an existing system.
The third task was more difficult than the first two. We describe each task below. 

\textbf{Task 1 (Mars Rover)} involved the implementation of the public API for moving a vehicle on a grid.
It was simple and algorithmic in nature, but required several edge cases to be handled as well as correctly parsing the input. The task is commonly used in the agile community to practice coding techniques. 
The requirements were described as six user stories, which were later refined to 11 sub-stories for acceptance testing. 
The participants were given the six user stories, the stub for the class definition, including the API signature (8 LOC), and a stub of the JUnit test class (9 LOC). 
The acceptance test suite had 89 JUnit assert statements that covered the 11 sub-stories.
We allocated four hours for this task.

\textbf{Task 2 (Bowling Scorekeeper)} required the implementation of a system to keep scores for a bowling game. 
This task is also well known and has been used in previous TDD studies \cite{George:2003:IIT}.

Also algorithmic in nature, its specification included 13 user stories to be implemented in a specific order.
Each user story contained an example scenario. 
The participants were given a stub containing two classes (23 and 28 LOC) with the signatures  of the methods to be implemented. A stubbed test class (9 LOC) was also provided.
The acceptance test suite of this task had 58 JUnit assert statements that covered the 13 user stories.
We also allocated four hours for this task.

\textbf{Task 3 (MusicPhone)} was about extending a concert recommendation system. 
The system has a traditional three-tier architecture that mimics a real-world application (see the task documentation included in the replication package \cite{Fucci2016}). 
We provided the participants with a partially working implementation consisting of 13 classes, and 4 interfaces (1033 LOC). 
The existing implementation included the UI and data access tiers. 
MusicPhone does not use any additional frameworks with the exception of JUnit for unit testing and Swing for the user interface. 
The existing part of the task was developed using the Singleton  pattern, which was explained to the subjects before the start of the experiment. 
The subjects were also given a cheat sheet of the architecture before they started working on the task. 
The participants were asked to implement three missing requirements in the business logic tier. They did not have to modify the presentation tier. 
These requirements were later refined into 11 sub-stories for acceptance testing.

In order to implement these requirements, the participants needed to understand the existing system and its architecture.
We provided the participants with a smoke test (38 LOC, 6 JUnit assertions) as an example of the usage of the existing classes (how to instantiate them and how they interact).

The first requirement was algorithmic, and the remaining requirements were more structural in nature. 
The associated acceptance test suite consisted of 132 JUnit assert statements, covering the 11 sub-stories. 
We allocated five hours for this task because it was deemed more difficult than Task 1 and Task 2.

Table \ref{tbl:tasks} shows how the participants were distributed over the tasks during the four runs of the workshop. 
\begin{table*}[!ht]
\centering
\caption{Allocation of the participants to tasks and techniques (ITL indicates iterative test-last).}
\begin{tabular}{ccc|cc|cc}
\hline
     Company     & \multicolumn{2}{c}{Task 1}      & \multicolumn{2}{|c}{Task 2}      & \multicolumn{2}{|c}{Task 3}      \\ \hline
Company A (3 runs) & \multicolumn{2}{c}{18 (ITL)} & \multicolumn{2}{|c}{19 (TDD)} & \multicolumn{2}{|c}{21 (TDD)} \\ 
Company B (1 run) & 6 (ITL)       & 9 (TDD)      &  2 (ITL)       & 8 (TDD)      & \multicolumn{2}{c}{16 (TDD)}      \\ \hline
\end{tabular}
\label{tbl:tasks}
\end{table*}

The total number of data points collected over the three tasks and three workshop runs was 99. After removing the data points containing missing values, our final dataset consisted of 82 data points (27 for Task 1, 25 for Task 2, and 30 for Task 3).

\subsection{Instrumentation}\label{sec:tool}
The development environment used by the participants included Java version 6, Eclipse 4, and JUnit 4.
The IDE was instrumented with the Besouro tool \cite{Becker:2014hz} to collect data about the development cycles. 

Besouro's data were used to calculate the metrics representing the factors. This tool is capable of recognizing development actions (e.g., creating a test class, running the unit tests, modifying a method) taking place inside the development environment and logging them along with a timestamp. 
The information is then used to aggregate a sequence of development actions into identifiable micro-cycles. 
The cycles are then classified as test-first compliant, refactoring, or non-test-first compliant (with three subcategories: test-last production, test addition, regression test) according to an expert system based on the heuristics presented in Kou et al. \cite{Kou:2010kv}.
This classification gives rise to the type of the cycle used in defining the factors $SEQ$ and $REF$. 
For example, the following sequence of development actions is classified as a test-last production cycle since the production code was written (and compiled) before the unit test:
\begin{framed}
\footnotesize
\begin{itemize}
	\item \textit{EditAction 1397055646 Frame.java ADD int score() METHOD bytes: 127}
	\item \textit{CompilationAction 1397055689 BowlingGame OK}
	\item \textit{EditAction 1397055691 FrameTest.java ADD void testScore() METHOD bytes: 391}
	\item \textit{UnitTestSessionAction 1397055770 FrameTest OK}\\
\end{itemize}
\end{framed}
As another example, the following sequence is classified as test-first compliant, or a test-first production cycle:
\begin{framed}
\footnotesize
\begin{itemize}
	\item \textit{EditAction 1397056134 BowlingGameTest.java ADD void testScore() METHOD bytes: 126}
	\item \textit{UnitTestSessionAction 1397056135 BowlingGameTest FAIL}
	\item \textit{FileOpenedAction 1397056152 BowlingGame.java}
	\item \textit{EditAction 1397056173 BowlingGame.java ADD int score() METHOD bytes: 340}
	\item \textit{UnitTestSessionAction 1397056176 BowlingGameTest OK}	
\end{itemize}
\end{framed}
In this last example, a unit test for a not-yet-existing functionality is written first. The test fails since there is no production code that can be exercised by it. The developer then writes some production code after which the test passes.

Each development cycle also has a duration derived from the timestamps (not shown here) of the delimiting actions. The duration is used to calculate $GRA$ and $UNI$. 

\subsection{Subjects}
The subjects were software professionals of varying levels of experience. 
For the workshops organized at Company A's sites,  24 developers signed up, but 22 participated in total. 
In Company B, after 19 initial signups, 17 participated in the workshop. 
All the subjects except two had at least a bachelor's degree in computer science or a related field. 
The average professional experience in Java development was 7.3 years (standard dev. = 4.1, min. = 1, max. = 15 years). 
Before the workshop, the participants were asked to state their level of skill in Java and unit testing on a 4-point Likert scale. 
Table \ref{tbl:skills} shows the breakdown of their answers. 

\begin{table}[!ht]
\centering
\scriptsize
\caption{Distribution of the subjects over the four skill levels for Java and unit testing (in percentage).}
\label{tbl:skills}
\begin{tabular}{ccccccc}
\cline{3-6}
\multicolumn{1}{l}{}                        & \multicolumn{1}{c}{}              & \multicolumn{4}{c}{Unit testing}                     \\ \cline{3-6} 
\multicolumn{1}{c}{}                        &                                   & None & Novice & Intermediate & Expert \\ \cline{3-6} 
\multicolumn{1}{|c|}{\multirow{4}{*}{\rotatebox[origin=c]{90}{Java}}} & \multicolumn{1}{l|}{None}         & 2.5  & 9.5    & 2.5          & 2.5         \\
\multicolumn{1}{|c|}{}                      & \multicolumn{1}{l|}{Novice}       & 27   & 17     & 9.5          & 0           \\
\multicolumn{1}{|c|}{}                      & \multicolumn{1}{l|}{Intermediate} & 2.5  & 9.5    & 2.5          & 5           \\
\multicolumn{1}{|c|}{}                      & \multicolumn{1}{l|}{Expert}  & 0    & 2.5    & 2.5          & 5          
\end{tabular}
\end{table}

Most of the subjects had basic experience with Java and unit testing. 
Few subjects (5\%) considered themselves experts in both Java and unit testing. Fewer (2.5\%) said they had neither skill. No one declared to be expert unit testers.
To sum up, the subject sample was composed  mostly of mid-level Java developers with basic unit testing skills, but experience and skill still varied. In order to establish a better baseline and level the playing field, the subjects were given a five-hour hands-on tutorial on unit testing principles using JUnit and a five-hour training on applying TDD in different contexts. 

\subsection{Data Analysis Techniques}\label{sec:techniques}
\begin{table*}[!ht] \centering 
  \caption{Descriptive statistics for the dataset used in the study (n=82).} 

\begin{tabular}{@{\extracolsep{5pt}}lcccccccc} 
\\[-1.8ex]\hline 
\hline \\[-1.8ex] 
Statistic & \multicolumn{1}{c}{Mean} & \multicolumn{1}{c}{St. Dev.} & \multicolumn{1}{c}{Min} & \multicolumn{1}{c}{Pctl(25)} & \multicolumn{1}{c}{Median} & \multicolumn{1}{c}{Pctl(75)} & \multicolumn{1}{c}{Max} \\ 
\hline \\[-1.8ex]  
QLTY  & 65.55 & 19.25 & 0.00 & 50.00 & 66.67 & 81.67 & 100.00 \\ 
PROD & .24 & .26 & .00 & .05 & .12 & .35 & .86 \\
GRA  & 7.93 & 9.53 & .87 & 2.78 & 4.36 & 8.95 & 48.97 \\ 
UNI & 4.52 & 9.14 & .00 & 2.43 & 4.52 & 8.17 & 51.32 \\ 
SEQ  & 32.79 & 21.86 & .00 & 16.66 & 33.33 & 47.33 & 87.50 \\ 
REF & 24.39 & 18.50 & .00 & 8.71 & 23.07 & 34.52 & 71.42 \\ 
\hline \\[-1.8ex] 
\end{tabular} 
  \label{tbl:descriptive} 
\end{table*} 
Our analysis starts with the characterization of the data with descriptive statistics. 
The next step is to look for correlations among the variables. 
This allows us to evaluate the redundant information in the factors (independent variables) and get a glimpse of the associations between the outcome (dependent) variables and factors.
Next we create multiple linear regression models to identify the relationship between the outcome variables and the complete set of factors.
Finally, we use feature selection to create the model that includes only the most salient factors.
For each model we report the effect size as adjusted $R^2$, which indicates the proportion of the variance in the outcome variable that the factors are able to explain. Thus $1 - \mathrm{adjusted \ }R^2$ can be attributed to unknown factors, or inherent variability in the outcome variable. 
As opposed to the p-value, which informs whether the association is present by chance or not, the effect size informs about the \textit{magnitude} of the association.
The soundness of the models is checked by applying the proper assumption diagnostics. 
The data analysis was carried out using R version 3.1.2. with $\alpha$ level set to .05.
The dataset used in the analysis is included in the replication package \cite{Fucci2016}.
\section{Results}\label{sec:analysis}

We collected six measures. The main four process dimensions ($GRA$, $UNI$, $SEQ$, and $REF$) constitute the factors (independent variables). 
$QLTY$, and $PROD$ are the outcome (dependent) variables.

The dataset includes 82 complete observations. An observation was considered complete when there were no missing values for any of the variables over the three tasks.
\subsection{Descriptive Statistics}\label{sec:descriptive}
\begin{figure*}
        \centering
        \begin{subfigure}[b]{0.49\textwidth}
                \includegraphics[width=\textwidth]{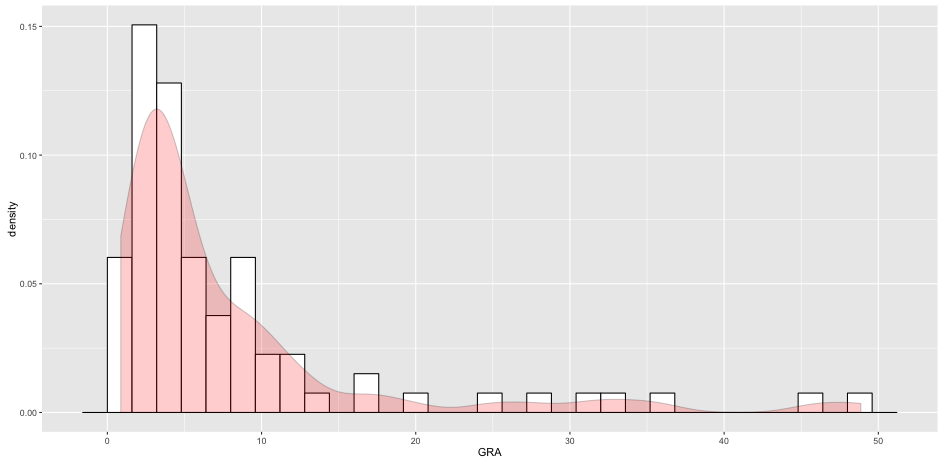}
                \caption{Granularity}
                \label{fig:gra_hist}
        \end{subfigure}%
        ~ 
        \begin{subfigure}[b]{0.49\textwidth}
                \includegraphics[width=\textwidth]{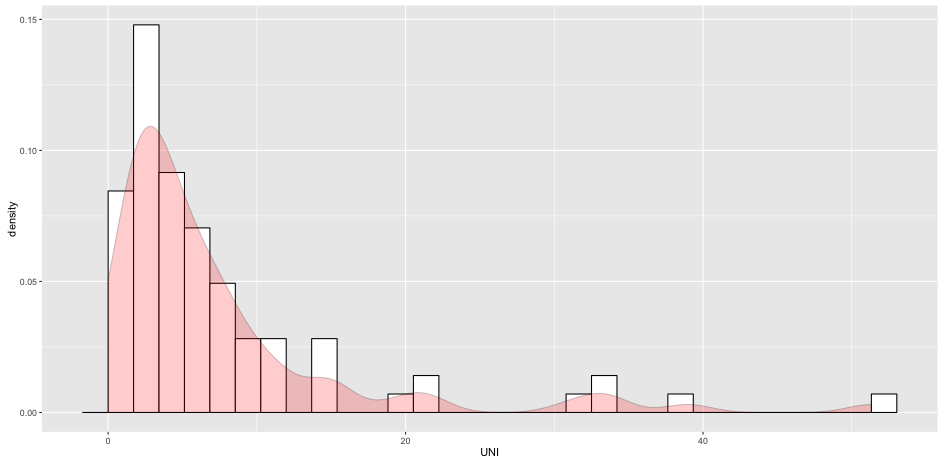}
                \caption{Uniformity}
                \label{fig:uni_hist}
        \end{subfigure}
        ~ 
        \begin{subfigure}[b]{0.49\textwidth}
                \includegraphics[width=\textwidth]{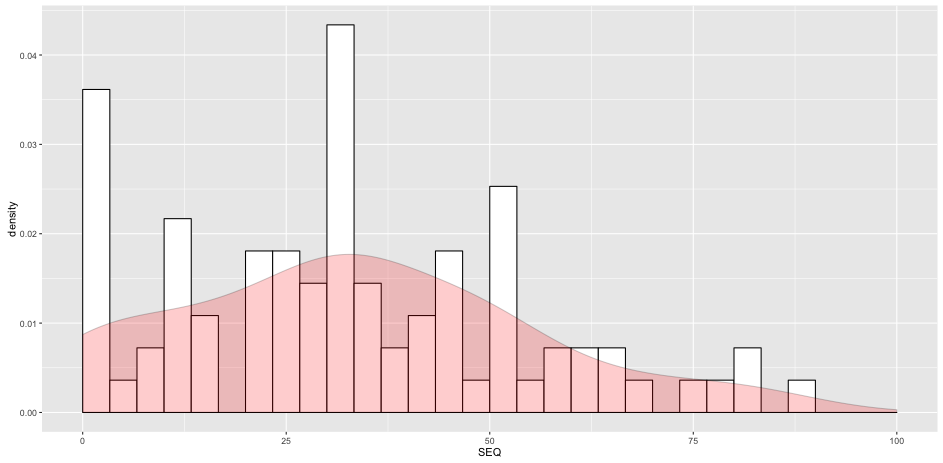}
                \caption{Sequencing}
                \label{fig:seq_hist}
        \end{subfigure}
         \begin{subfigure}[b]{0.49\textwidth}
                \includegraphics[width=\textwidth]{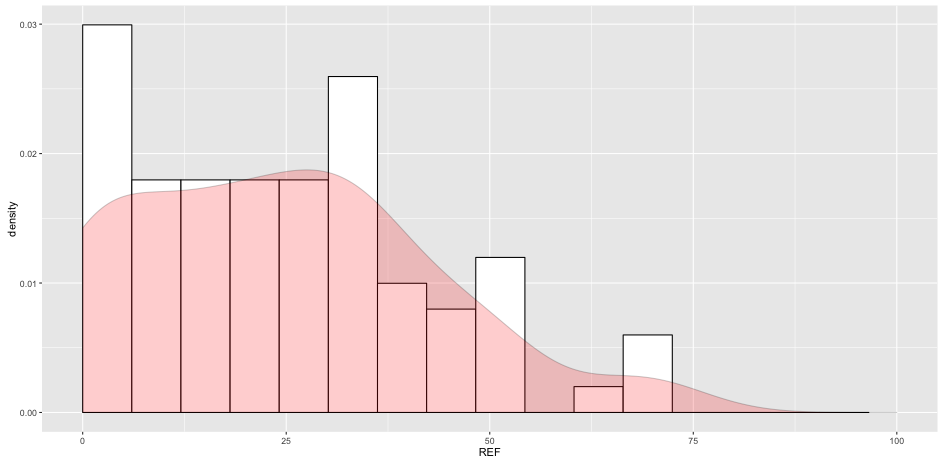}
                \caption{Refactoring}
                \label{fig:ref_hist}
        \end{subfigure}
         \begin{subfigure}[b]{0.49\textwidth}
                \includegraphics[width=\textwidth]{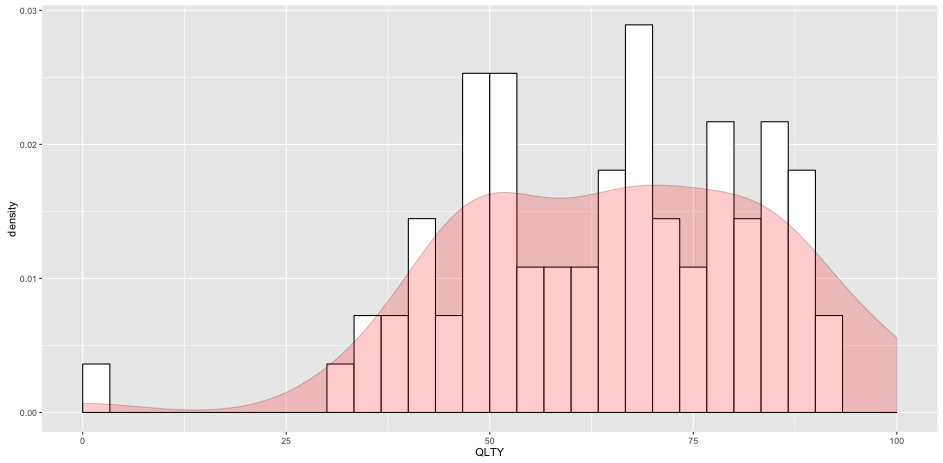}
                \caption{Quality}
                \label{fig:qlty_hist}
        \end{subfigure}%
         \begin{subfigure}[b]{0.49\textwidth}
                \includegraphics[width=\textwidth]{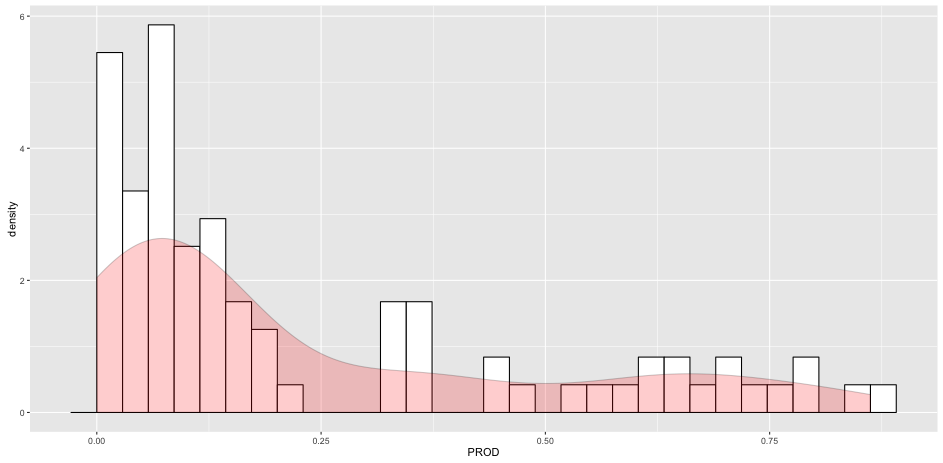}
                \caption{Productivity}
                \label{fig:prod_hist}
        \end{subfigure}%
        \caption{Histograms and density curve for the study variables. The x-axis for the histograms of granularity and uniformity are cut to maximum value of the dataset, instead of extending to the theoretical maximum.}
            \label{fig:hists}
\end{figure*}

The descriptive statistics for the dataset are reported in Table \ref{tbl:descriptive}.
The histogram and probability density of each variable is shown in Figure \ref{fig:hists}. 

$QLTY$ appears normally distributed, with about 75\% of the observations in the 50-100 range, as shown in Figure \ref{fig:qlty_hist}. 
$PROD$ appears positively skewed and multi-modal as shown in Figure \ref{fig:prod_hist}. Approximately 25\% of the observations are in the 0-0.25 range (min = 0, max = .86). 
The shapes of the distribution of $QLTY$ and $PROD$ are consistent with those of a previous experiment conducted with student subjects \cite{Fucci:2013wx}.

The distributions of $GRA$ (Figure \ref{fig:gra_hist}) and $UNI$ (Figure \ref{fig:uni_hist}), variables capturing the cyclical characteristics of the process, are positively skewed. 
For both, the majority of the observations are in the 0-10 interval. The peek of the granularity distribution is close to the suggested five minutes; also the peek and skewness of the uniformity distribution suggest that the majority of the subjects kept a consistent rhythm ($UNI$ = 0 minute indicates that all the cycles had the same duration).

Figure \ref{fig:seq_hist} shows that some subjects, around a quarter, wrote tests first only to some extent (approximately one-third of the total cycles completed). Subjects in the upper quantile wrote test-first approximately half of the time. 
The distribution of $REF$ is positively skewed, as shown in Figure \ref{fig:ref_hist}. 
The subjects applied refactorings approximately one-third of the time on average, but a quarter of them refactored less than 10\% of the time.
This behavior is not surprising as it often happens in real projects, as was observed by Aniche et al. \cite{Aniche2010Most}. 
%

\subsection{Correlation Analysis}\label{sec:corrs}
In this section we check for significant correlations between the variables, using the Spearman $\rho$ coefficient \cite{McCrumGardner:2008ee}. 
Spearman's correlation is a rank-based correlation measure that does not rest upon an assumption of normality, and it is robust with respect to outliers.

Table \ref{tbl:corrs} summarizes the correlations among the four factors.
The factors dealing with the cyclical characteristics of the process, $GRA$ and $UNI$, are positively correlated ($\rho$ = .49, p-value = .0001). The shorter the cycles tend to be, the more uniform they are, and vice versa. 
Another significant correlation exists between $GRA$ and $REF$, but the magnitude is small ($\rho$ = $-$.22, p-value = .02). Coarser processes appear to involve reduced refactoring effort, and vice versa.

Table \ref{tbl:corrs2} summarizes the correlations among the dependent (outcome) variables and independent variables (factors).
$QLTY$ is negatively correlated with $GRA$, i.e., the smaller the cycle, the higher the quality, and vice versa.  The correlation is significant ($\rho=-.25$, p-value=.02) and its magnitude is medium \cite{ellis2010essential}. 
A similar negative association exists between $QLTY$ and $UNI$; the more uniform the development cycles, the better the external quality ($\rho=-.36$, p-value = .01), and vice versa.

Unlike $QLTY$, none of the factors correlate significantly with the outcome variable $PROD$.

\begin{table}[!ht]
\centering
\caption{Correlations among the factors. \\ The values above the diagonal represent Spearman's correlation coefficient; the values below represent the p-value.}
\begin{tabular}{lllll}
\hline
    & GRA     & UNI & SEQ   & REF   \\ \hline
GRA &         & .49 & .11  & -.22  \\
UNI & .0001 &     & -.11 & .01 \\
SEQ & .15     & .92 &       & -.48 \\
REF & .02     & .30 & .10   &     \\
\hline 
\end{tabular}
\label{tbl:corrs}
\end{table}

\begin{table}[!ht]
\centering
\caption{Correlations between the factors and outcomes. 
 The p-values are reported in parentheses.}
\begin{tabular}{lllll}
\hline
&GRA& UNI& SEQ& REF\\ \hline
QLTY & \begin{tabular}[c]{@{}l@{}}-.27\\ (.02)\end{tabular} & \begin{tabular}[c]{@{}l@{}}-.36 \\ (.01)\end{tabular} & \begin{tabular}[c]{@{}l@{}}.12\\ (.29)\end{tabular} & \begin{tabular}[c]{@{}l@{}}-.11\\ (.31)\end{tabular}       \\
PROD & \begin{tabular}[c]{@{}l@{}}-.20\\ (.07)\end{tabular} & \begin{tabular}[c]{@{}l@{}}-.25\\ (.20)\end{tabular} & \begin{tabular}[c]{@{}l@{}}.03\\ (.81)\end{tabular} & \begin{tabular}[c]{@{}l@{}}-.17\\ (.11)\end{tabular}\\
\hline
\end{tabular}
\label{tbl:corrs2}
\end{table}

\subsection{Regression Models}\label{sec:multivariate}
In this section, we explore the relationships between the four factors and two outcome variables and find optimal models that explain how useful the factors are in predicting the observed outcomes.
We start with a full model that includes all the factors and employ a backward elimination approach to find the model that minimizes information loss according to Akaike's Information Criterion (AIC) \cite{burnham2004multimodel}. 
Minimizing AIC is preferred to maximizing adjusted $R^2$ because it favors the simplest model \cite{burnham2004multimodel}, whereas adjusted $R^2$ can be biased towards the more complicated model.
The initial model we propose has two high-level components, with  possible interaction within the two high-level components, but no interactions across them.
\begin{itemize}
	\item The high-level component $CYCLE$ deals with the cyclical characterisics of the development process. This component includes the factors granularity and uniformity, and the interaction between them. 
	\item The high-level component $TYPE$ deals with the specific approach used within cycles. This component includes the factors sequencing and refactoring, and the interaction between them.
\end{itemize}
It is reasonable to assume that $CYCLE$ and $TYPE$ do not interact because they represent orthogonal higher-order concepts.

Hence, our model is $CYCLE + TYPE$, where:
\begin{itemize}
	\item $CYCLE = GRA + UNI + (GRA:UNI)$
	\item $TYPE = SEQ + REF + (SEQ:REF)$
\end{itemize}
Here the  symbol ":" denotes interaction between two factors. Thus we have:
\begin{equation}\label{eq:model}
\begin{split}
	OUTCOME \sim GRA + UNI + GRA:UNI \ \ + \\ SEQ + REF + SEQ:REF
\end{split}
\end{equation}
where $OUTCOME \in$ [$QLTY$, $PROD$].

The results for this model are reported in Table \ref{tbl:multiv_qlty}.
The model in Equation \ref{eq:model} is statistically significant for both outcome variables according to F-statistic. 
The observed effect sizes (measured by adjusted $R^2$) are considered \textit{small-medium} \cite{ellis2010essential}, typical for  studies involving human factors \cite{cohen1992power,kampenes2007systematic}. 
However, the factor coefficients and interaction terms are all statistically insignificant at the given $\alpha$-level. 
The correlations observed between the two factors $GRA$ and $UNI$ (see Section \ref{sec:corrs}) may indicate a problem of multicollinearity in the models. 
Although it does not impact the power of the model, multicollinearity\footnote{Multicollinearity results from two or more factors in a model being significantly correlated, resulting in redundancy, and possible overfitting.} has the risk of inaccurate regression coefficients and p-values \cite{farrar1967multicollinearity}.
We can assess the extent of such overfitting using the Variance Inflation Factor (VIF).
VIF indicates to what extent the variances of the coefficients estimates are inflated due to multicollinearity. 
A VIF value exceeding 4.0 is interpreted as problematic multicollinearity \cite{Jobson:1999:multivariate}. 
The interaction term $GRA:UNI$ exhibits a VIF of 5.53, which is excessive. For the factors and the other interaction term, we have  $\mathrm{VIF}(GRA) = 3.39$, $\mathrm{VIF}(UNI) = 1.91$, $\mathrm{VIF}(SEQ) = 2.38$, $\mathrm{VIF}(REF) = 3.24$, and $\mathrm{VIF}(SEQ:REF) = 2.56$, which are acceptable.   

\begin{table*}[!htbp] \centering 
  \caption{Regression summary for the models in Equation \ref{eq:model}. Standard error reported in parentheses.} 
\begin{tabular}{@{\extracolsep{5pt}}lcc} 
\\[-1.8ex]\hline 
\hline \\[-1.8ex] 
 & \multicolumn{2}{c}{\textit{Outcome variable:}} \\ 
\cline{2-3} 
\\[-1.8ex] & QLTY & PROD \\ 
\hline \\[-1.8ex] 
 GRA & $-$.36 (0.19)  & $-5.03e{-}03$  ($2.58e{-}02$) \\ 
  & p-value = .65  & p-value =  .05* \\  
  & & \\
 UNI & $-$.62  (.49)  & $-9.96e{-}03$ ($6.53e{-}03$) \\ 
  & p-value = .20 & p-value =  .13 \\  
  & & \\
 SEQ & .002 (.14)  & $-1.48e{-}03$ ($1.80e{-}03$) \\ 
  & p-value = .86 & p-value =  .41 \\  
  & & \\
 REF & $-$0.26 (.20) & $-4.91e{-}03$ ($2.53e{-}03$)\\ 
  & p-value =  .17 & p-value =  .05 \\ 
  & & \\
 GRA:UNI & .006 (.04) & $1.57e{-}04$  ($1.87e{-}04$) \\ 
  & p-value =  .65  & p-value =  .40 \\  
  & & \\
 SEQ:REF & .001 (.005)  & $3.60e{-}05$ ($7.49e{-}05$) \\ 
  & p-value =  .80  & p-value =  .62 \\ 
  & & \\
 Constant & 77.27 (7.14) & 0.47 (.09)\\ 
  & p-value =  .000  & p-value =  .000 \\ 
\hline \\[-1.8ex] 
Adjusted R$^{2}$ [conf. int.] & .09 [.01, .26]&  .08 [.01, .26] \\ 
Residual Std. Error (df = 75) & 18.40 & .24 \\ 
F Statistic (df = 6; 75) & 2.42  & 2.72 \\ 
p-value & .03 & .04 \\
\hline 
\hline \\[-1.8ex] 
\end{tabular} 
  \label{tbl:multiv_qlty} 
\end{table*}  

Based on these results, we drop the problematic interaction factor $GRA:UNI$ from further analysis.
Therefore, we modify the candidate model for $QLTY$ to be:
\begin{equation}\label{eq:model_multicoll}
	QLTY \sim GRA + UNI + SEQ + REF + SEQ:REF 
\end{equation}

\subsubsection{Feature Selection}
We now investigate whether the model in Equation \ref{eq:model_multicoll} can be further simplified by applying feature selection. 
Our approach, stepwise backward selection \cite{1100705}, iteratively drops the least significant variable to form a new model based on its AIC. 
The re-inclusion of any previously dropped variables is considered at the end of each iteration. 

AIC represents the information loss for a given model relative to the theoretical ideal, unlike the Chi-squared goodness-of-fit approaches in which the absolute model fit is assessed \cite{Burnham2003Model}.
AIC takes into account the complexity of the model, i.e., the number of variables. 
As opposed to $R^2$-based indicators \cite{Burnham2003Model}, it does not inflate fitness after more variables have been added.
Another relative fitness index that takes into account model parsimony similar to AIC is BIC \cite{burnham2004multimodel}. 
BIC tends to penalize the number of variables in the model more than AIC. 
However, unlike BIC, AIC reaches an asymptotically optimal model under the assumption that the best theoretical model is not among the possible candidates \cite{yang2005can}. This property of AIC makes it favorable to BIC in our case: 
the best model for $QLTY$ and $PROD$ cannot be constructed using only the factors measured in this study, and therefore we use AIC as the basis for the selection criterion. We are looking for the model with the minimum AIC\footnote{The theoretical best value for AIC is minus infinity.}. 

Applying backwards selection, after three iterations we obtain the models presented in Table \ref{tbl:QLTYb} for $QLTY$ and Table \ref{tbl:PRODb} for $PROD$. 
These two models represent the best compromise between goodness of fit and simplicity with minimum loss of information.
\begin{table}[!htbp] \centering 
  \caption{Regression summary for the $QLTY$ model having the best AIC (483.50). Standard error reported in parentheses.} 
\begin{tabular}{@{\extracolsep{5pt}}lc} 
\\[-1.8ex]\hline 
\hline \\[-1.8ex] 
 & \multicolumn{1}{c}{\textit{Outcome variable:}} \\ 
\cline{2-2} 
\\[-1.8ex] & QLTY \\ 
\hline \\[-1.8ex] 
 GRA & $-$.34 (0.17) \\ 
  & p-value =  .04 \\ 
   & \\
  UNI & $-$.43 (.25) \\ 
  & p-value =  .09 \\ 
   & \\
  REF & $-$.25 (.11) \\ 
  & p-value =  .03 \\ 
   & \\
  Constant & 77.70 (4.05) \\ 
  & p-value =  .000 \\ 
 \hline \\[-1.8ex] 
Adjusted R$^{2}$ [conf. int.] & .12 [.01, .27]\\ 
Residual Std. Error & 17.98 (df = 78) \\ 
F Statistic & 4.86 (df = 3; 78) \\ 
p-value & .003 \\
\hline 
\hline \\[-1.8ex] 
\end{tabular} 
\label{tbl:QLTYb}
\end{table}


\begin{table}[!htbp] \centering 
  \caption{Regression summary for the $PROD$ model having the best AIC (-239.30). Standard error reported in parentheses.} 
\begin{tabular}{@{\extracolsep{5pt}}lc} 
\\[-1.8ex]\hline 
\hline \\[-1.8ex] 
 & \multicolumn{1}{c}{\textit{Outcome variable:}} \\ 
\cline{2-2} 
\\[-1.8ex] & PROD \\ 
\hline \\[-1.8ex] 
 GRA & $-$.003 (.002) \\ 
  & p-value =  .11 \\ 
   & \\
  UNI & $-$.003 (.002) \\ 
  & p-value =  .08 \\ 
   & \\
  REF & $-$.003 (.001) \\ 
  & p-value =  .02 \\ 
   & \\
  Constant & .39 (.05) \\ 
  & p-value =  .000 \\ 
 \hline \\[-1.8ex]  
Adjusted R$^{2}$ & .10 [.01, .28] \\ 
Residual Std. Error & .22 (df = 78) \\ 
F Statistic & 4.14 (df = 2; 78) \\ 
p-value & .008 \\
\hline 
\hline \\[-1.8ex] 
\end{tabular} 
\label{tbl:PRODb}
\end{table} 

Both models are significant and exhibit an adjusted $R^2$ indicating a \textit{medium} effect size. 
Notice that $SEQ$ and its interaction with $REF$ ($SEQ$:$REF$) have been dropped from both models. This is surprising as it implies the sequence in which writing test and production code are interleaved is not a prominent feature. The finding counters the common folklore within the agile software development community.

The factors related to the higher-level $CYCLE$ component ($GRA$ and $UNI$) as well as remainin $TYPE$ component (refactoring effort, $REF$) are negatively associated with both outcomes. In the end, the models for $QLTY$ and $PROD$ involve exactly the same factors. 


That refactoring has a negative relationship with both outcome variables seems counterintuitive.
One reason may lie in the metric $REF$ used to measure refactoring effort.  
$REF$ may not reliably measure the construct that it targets, which raises a construct validity threat. This has to do with the measure's inability to  differentiate between useful and harmful variants of refactoring. 
Most cycles detected as refactoring cycles could have been associated with \textit{floss refactoring} \cite{MurphyHill:2012ca}, a practice that mixes refactoring with other activities. 
A typical example involves sneaking in production code that implements new functionality while refactoring. The developer is peforming refactoring, realizes that a piece of functionality is missing and mixes in the proper production code, but the code is not covered by any tests, possibly causing a feature interaction and introducing a hidden bug. 
Floss refactoring is believed to be potentially harmful, negatively affecting quality and productivity. Pure refactoring cycles are difficult to detect accurately without before-and-after code coverage comparison, which the process instrumentation tool we used did not perform.

\subsubsection{Assumption Diagnostics}
We validate the assumptions for the models by running the regression diagnostic proposed by Pena and Slate \cite{pena2006global}.
\begin{table}[!ht]
\centering
\caption{Multiple regression diagnostics for $QLTY$ model in Table \ref{tbl:QLTYb}.}
\begin{tabular}{lrrl}
  \hline
 & Value & p-value & Decision \\ 
  \hline
  Skewness & 1.93 & .16 &  Acceptable. \\ 
  Kurtosis & .20 & .65 &  Acceptable. \\ 
  Link Function & 1.63 & .26 &  Acceptable. \\ 
  Heteroscedasticity & .10 & .74 &  Acceptable. \\ 
  \hline
\end{tabular}
\label{tbl:QLTYd}
\end{table}


\begin{table}[!ht]
\centering
\caption{Regression diagnostics for $PROD$ model in Table \ref{tbl:PRODb}.}
\begin{tabular}{lrrl}
  \hline
 & Value & p-value & Decision \\ 
  \hline
  Skewness & 7.21 & .07 &  Acceptable. \\ 
  Kurtosis & .81 & .36 &  Acceptable. \\ 
  Link Function & 1.49 & .22 & Acceptable \\ 
  Heteroscedasticity & 2.26 & .13 &  Acceptable. \\ 
  \hline
\end{tabular}
\label{tbl:PRODd}
\end{table}
\begin{figure*}[ht]
        \centering
        \includegraphics[width=\textwidth]{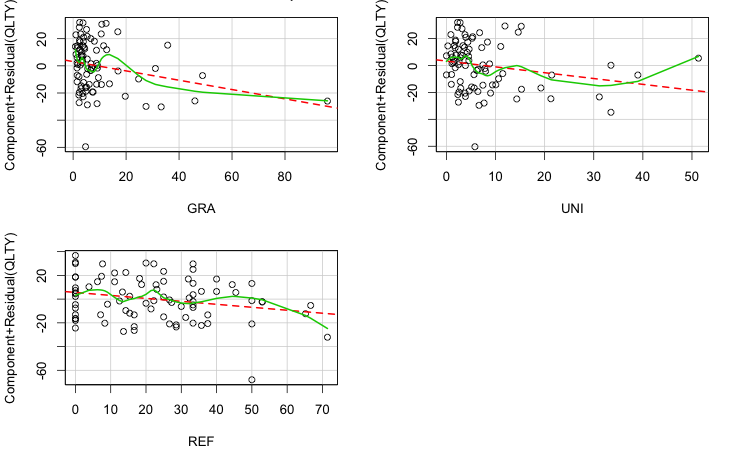}
        \caption{CCPR plots for model in Table \ref{tbl:QLTYb}}
        \label{fig:cr_qlty}
\end{figure*}

\begin{figure*}[ht]
        \centering
        \includegraphics[width=\textwidth]{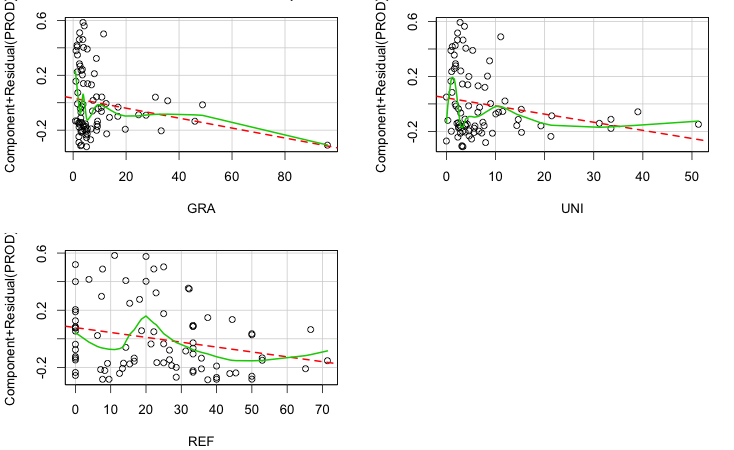}
        \caption{CCPR plots for model in Table \ref{tbl:PRODb}}
        \label{fig:cr_prod}
\end{figure*}
As reported in Tables \ref{tbl:QLTYd} and  \ref{tbl:PRODd}, the models overall meet their statistical assumptions (p-value $>$ $\alpha$-level).  
When a model includes several factors, partial residuals plots are helpful in visualizing the relationships between the factors and the outcome variable one at a time.
Figures \ref{fig:cr_qlty} and \ref{fig:cr_prod} show the Component and Component+Residuals (CCPR) plots for the $QLTY$ and $PROD$ models, respectively. 
The CCPR plot is a variation of a partial residual plot where the outcome values and the partial residuals are plotted against a chosen factor.
The dashed line in each plot represents the fitted linear model for the corresponding factor. This is the component line.
The circles represent the partial residuals. The solid curve is fitted to the residuals using the LOESS method---a standard, nonparametric method to fit a curve to a set of observations based on locally weighted polynomial regression \cite{cleveland1988locally}.
A systematic departure of the residuals curve from the component line indicates a poor fit \cite{xia2009model}, which does not appear to be the case in Figures \ref{fig:cr_qlty} and \ref{fig:cr_prod}. We observe sporadic local departures of the LOESS curves from the component lines, especially at some extremes, but the curves do not "run away" persistently.   

\subsubsection{Interpretation of Models}
We briefly interpret the significant coefficients in the two models.

Regarding the $GRA$ coefficient in $QLTY$ (Table \ref{tbl:QLTYb}), reducing the cycle length further from the often suggested values of five to ten minutes results in little improvement. 
The improvement can be as high as 14\% when cycle length is reduced from its maximum observed value at around 50 minutes down to the average observation of eight minutes in the nominal range of the dataset. 

With respect to $REF$ coefficient, we have similar observations, however, we have to qualify them with the possibility that the detected refactoring cycles might have been dominated by the harmful variety (further discussed in the next section), hence the negative coefficient. 
Limiting refactoring effort from the average level (23\%) to its first quartile mid-point (9\%), and similarly from the third quartile midpoint to the average level, results in less than 4\% improvement in external quality. However, a reduction from the extreme case (70\%) to the average case results in about 12\% improvement. 

In the $PROD$ model (Table \ref{tbl:PRODb}), only $REF$ has a significant coefficient, but again it is subject to the construct caveat of being possibly overrepresented by the harmful variety of refactoring.
Keeping this qualification in mind, doing less refactoring is associated with an increase in productivity. 
This observation is not surprising since spending more time on refactoring should imply spending less time on adding new features, thus reducing productivity. 
In particular reducing refactoring effort from 70\% of the time (the maximum observed value) to the average case in the dataset---a reduction of $\sim$45\%---results in a productivity improvement of 18\%.
In summary, the final optimal models include the factors for granularity, uniformity, and refactoring effort, but not sequencing. They also omit interactions between these factors.
However, at this point the role of uniformity vis-\'a-vis external quality and of granularity vis-\'a-vis productivity are still uncertain since the corresponding coefficients are insignificant in the underlying models, although the models  themselves are significant. 
\section{Threats to Validity}\label{sec:ttv}
In this section, we discuss the threats to the validity applicable to our study  following the classification by Wohlin et al. \cite{wohlin2000experimentation}: 
\begin{enumerate}
  \item Internal validity: threats that can lead to a misinterpretation of the relationship between independent and dependent variables.
  \item External validity: threats that can limit the applicability of the results to a greater extent, e.g., in industrial practice.
  \item Construct validity: threats that relate to how faithfully the measures capture the theoretical constructs and how suitably the constructs are operationalized to preserve their integrity.
  \item Conclusion validity: threats that concern the soundness of the statistical inferences made.
\end{enumerate}

The threats to the validity are presented in decreasing order of priority following the recommendations of Wohlin et al.  \cite{wohlin2000experimentation}.
We recognize that there are trade-offs between the different threats and focus on the types that are more important for \textit{applied research}.
Therefore, we give higher priority to the interpretation of the results and their applicability to relevant populations. 

\subsection{Internal Validity}
By the nature of its design, this is a single-group study. Therefore, there is no notion of a control group.
The study is subject to single-group threats. 
The factors, or independent variables, represent dimensions of applying a particular type of development process. 
We are interested in how the variation in the strength of the property described by each dimension of the process affects the outcome measures.
The dimensions cannot be controlled, as they arise naturally from applying a taught technique on a best effort basis.
They are neither all absent nor all present, but rather present to varying extents in a continuum. 
This is the reason this study uses correlation and regression analysis rather than an experimental design with defined control and experimental groups.
We noted in the beginning that this study was embedded in a larger ongoing study structured as a quasi-experiment. 
However, the treatment and control groups, TDD and ITL, defined in that larger context, do not apply to this study, as we generalized the groups to a higher-level process in which TDD and ITL are specific,  idealized cases. 
We are interested in the elementary dimensions of that high-level process.
We acknowledge the limitations of our design in terms of causality. 
Our main goal was to identify the most salient of those dimensions in terms of their potential consequences.

History and maturation effects are not a concern for this study since they apply to all subjects equally.
Threats related to testing and statistical regression to the mean  are not applicable either due to the design.

We mitigated the instrumentation threat by using multiple tasks. Two of these tasks were selected from previous studies, and a more complex task was introduced to increase the variation in observations.
We needed observations to be as diverse as possible.

The drop-out rate of 10\%, together with 8\% incomplete or excluded observations, may have eliminated a relevant subgroup of subjects. This in turn may have skewed the dataset. 

A selection threat exists due to convenience sampling. 
Our sample was based on volunteers and company-selected subjects. 
Such subjects tend to have different motivations and incentives different from those of randomly selected subjects. 
In general, volunteers tend to be more enthusiastic and motivated.
The converse might apply to company-selected subjects, which in addition poses a social threat.
The skill and experience profiles of the two groups may have been skewed compared to a random sample. 
These design choices could have influenced the results.
However, we needed to accept the compromises in the selection process because recruiting professionals is inherently difficult, and when possible, often subject to selection bias \cite{Misirli:2014gk}.

\subsection{External Validity} 
The participants in the study do not represent the entire population of software developers. 
Specifically, their skills in unit testing was limited. 
Professional experience level spanned a rather narrow range skewed towards the medium- to high-experience end. 
These two considerations limit the generalizability of the results to sub-populations with profiles that differ from that of the study.

The tasks consisted of fixed user stories, all known upfront to the subjects. We acknowledge that this may not be the case in real world settings. When requirements may change or not well known, leading the cycles with tests can provide advantages that were not detectable in this study:  test-first can help to understand the requirements, discover new/missing requirements, and find and fix requirements bugs. 

Two of the three tasks used in the study were artificial and not particularly representative of a real-word situation for a professional software developer. 
This poses a threat to the generalizability of our results.
To mitigate this concern and increase realism, we included a third task, and the three tasks together provided sufficient diversity and balance between algorithmic and architectural problem solving.    
The resulting diversity moderated the threat to external validity.

To minimize the interaction between setting and study goals, the  physical setting was kept as natural as possible for the participants.
The subjects worked at their own workstations throughout the study. 

\subsection{Construct Validity}
The study constructs were well defined, except possibly for the refactoring dimension. 
The factors granularity, uniformity, and sequencing are based on straightforward, mechanical definitions, and were accordingly measured by automated tools with straightforward heuristics.
Refactoring effort is more special since the underlying construct is difficult to capture, or even define clearly.
It was measured using an approximation. 
In the future, its detection may be made more accurate by more sophisticated analysis of developer actions---for example through before-and-after code coverage analysis to better differentiate pure refactoring from the potentially harmful, mixed kind known as \textit{floss refactoring} \cite{Aniche2010Most}.
Since the corresponding factor, $REF$, appears in the final regression models, a construct threat arises. In particular, the role of refactoring in external quality deserves deeper investigation. 

The outcome variables were based on measures used in previous studies. 
We had no reason to doubt their ability to capture the underlying constructs in a reasonable way. 
The productivity measure was, by design, intended to capture short-term productivity. 
Therefore, given the short duration of the tasks, it is possible that the participants did not achieve sufficient maturity for robust measurement. 
This could suggest an interaction between maturity and construct validity.

The outcome variables exhibit mono-method bias. 
We used a single measure to capture each outcome construct. For example, the quality measure was based exclusively on functional correctness. Other important aspects of external quality, in particular usability and usefulness,  were disregarded. 
Not triangulating the result with multiple measures reduces the validity. 
We are still looking for alternative, objective, and reliable means of measuring these constructs. 

Finally, the factors we studied are directly associated with only two particular outcome constructs, external quality and developer productivity. The factors may also have an association with other important outcome constructs, such as  internal quality (maintainability and understandability) and ability to deal with volatile requirements. These were not  addressed. The factors' impact on these other constructs can be significant, and thus worth investigating.   

\subsection{Conclusion Validity}
Subject variability was desirable because this was a regression study. The participants were professionals from only two companies, therefore our sample was not very heterogeneous.  
There were few subjects who were experts with both relevant skills. 
The majority of the participants had to learn the required skills to different extents during the workshops, which  possibly further reduced variability in the sample. 

Skill and experience level are often confounding factors in experimental designs. Their role in an observational regression study with no controlled factors is however less important. 

We believe the application domain in which the participants worked (security and gaming) was not a concern either for conclusion validity. 
The tasks were equally new and different to all participants.

The assumptions of the statistical tests used were verified for all the models. 
In the final models, the effect sizes---represented by the adjusted $R^2$---fell in the medium range \cite{ellis2010essential}.
This level is acceptable in the study's context because human behavior can introduce a great deal of unexplained variation \cite{kampenes2007systematic} that may be impossible to account for.
We remark that over 85\% of the variation in the outcome variables was still unaccounted for. 
However, the study was not focused on those  factors; rather, the goal was to identify the most important ones among four specific candidates.
\section{Related Work}
\label{sec:background}
\begin{table*}[!ht]
\centering
\caption{Summary of secondary studies about TDD}
\begin{tabular}{lllll}
\hline
 Reference & \parbox[t]{1cm}{Primary\\ Studies} & \parbox[t]{2.5cm}{Result \\External Quality} & \parbox[t]{2cm}{Result \\Productivity} & Notes \\
 \hline
 \cite{turhan2010effective}              &         32        &    Some improvement           &      Inconclusive                      &    \parbox[t]{3cm}{Following TDD \\ not enforced}   \\ \hline
 \cite{Munir:2014bp}         &   41              &        \parbox[t]{3cm}{Improvement in \\ high-rigor,high-relevance}        &            Inconclusive                       &    \parbox[t]{3cm}{Inconclusive when\\ studies are considered together}  \\ \hline
 \cite{rafique2013effects}         & 27                &      \parbox[t]{3cm}{Small improvements, \\more accentuated in industry }         &   \parbox[t]{3cm}{Inconclusive,\\ with a drop in industry} &                \parbox[t]{3cm}{Task size \\is a significant moderator}  \\   \hline 
\end{tabular}
\label{tbl:secondary}
\end{table*}
The effects of TDD have been extensively studied. The evidence accumulated is synthesized in several systematic literature reviews (Table \ref{tbl:secondary}). 

Turhan et al. \cite{turhan2010effective} reviewed controlled experiments, quasi-experiments and case studies of different scales in both industry and academia. Their results suggest that TDD overall has a moderate positive effect on external quality, although this result becomes weaker when only the most rigorous studies are considered. 
The results are contradictory for productivity overall, with controlled experiments exhibiting a weak positive effect.

A meta-analysis by Rafique and Misic \cite{rafique2013effects} reports that TDD improves external quality when compared to a waterfall approach. 
However, this improvement is not strong. Further, TDD becomes disadvantageous for the subset containing only academic studies in which it is compared to an iterative, test-last (ITL) process instead of a waterfall approach. 
This result suggests that sequencing might have a negative effect on external quality, which we haven't observed.
Productivity results are more inconclusive in that the authors report a small  productivity hit for TDD when comparing TDD with waterfall (attributed to differences in testing effort, with a larger hit for the industrial subgroup), but the effect, even though still small, is reserved when ITL is compared with TDD. 
This latter observation suggests sequencing might have a positive influence on productivity, which we have not been able to confirm, either. 
Rafique and Misic recommend further investigation with longer-term studies with larger tasks to allow the effects to become more visible. 
They also emphasize the importance of process conformance in TDD, which we capture using the four dimensions.

According to a subsequent systematic literature review by Munir et al. \cite{Munir:2014bp}, TDD is beneficial for external quality when only high-quality studies are considered. 
However, this improvement disappears over the whole dataset, which includes studies of different levels of rigour. 
Again, the productivity results are inconclusive. 
The authors recommend future studies to focus on industrial settings and professionals, and our study is in that category.  

Causevic et al.'s review \cite{Causevic:2011hp} identified what they refer to as insufficient \textit{adherence to protocol}, or process conformance (i.e., the extent to which the steps of TDD are followed), as one of the factors leading to the abandonment of the practice. 
The authors' review of industrial case studies showed that, although developers deviate from the process in several situations, conformance to the TDD process is regarded as important.
The authors suggest that industrial investigations of TDD consider the underlying process, which is what we endeavored to do in our study.

Therefore, in general, there is a consensus that the notion of \textit{process} deserves a closer look. 
We agree with this view. However, often TDD researchers equate process with the test-first dynamic of TDD. 
We think that the idea of \textit{process conformance} \cite{Munir:2014bp} or \textit{adherence to protocol} \cite{causevic2012evaluation} is not only limited to the sequencing dimension. 
Indeed, by looking closer at the other dimensions, we discover that granularity and uniformity of the development cycles trumps sequencing, at least in terms of external quality, and developer productivity.
In fact, in our previous studies, we investigated the relationship between sequencing and external quality and sequencing and productivity \cite{fucci2014impact, Fucci:2014bc} for TDD-like processes, including ITL. 
We did not find any evidence of such a relationship. 

M\"uller and H\"ofer \cite{muller2007effect} investigated the ways experts and novices differ in applying TDD, focusing on process conformance. 
To our knowledge, they are the only researchers who explicitly considered characteristics of TDD cycles other than sequencing. 
They used an automatic inference tool similar to Besouro \cite{Becker:2014hz}, which was used in our study. 
Similar to our approach, M\"uller and H\"ofer use this tool to log developer actions and roll them into cycles with different levels of compliance to a set of TDD rules. 
Remarkably, they found that average cycle length, close to our definition of granularity, and the cycle length's variation, close to our definition of uniformity, were the most important differentiators between experts and novices. 
In our study, we found a large and highly significant correlation between granularity and uniformity, suggesting that these two dimensions might capture the same information. 

Pancur et al. \cite{Pancur:2011ef} carried out two controlled experiments comparing TDD to ITL, while making sure that the process granularity of two treatments were comparable. 
Their rationale was to eliminate the possible effect of the granularity, thereby isolating the effect of sequencing. 
Again, they could not produce any evidence that sequencing affects productivity or external quality. 

Pancur et al. thus speculate that studies showing the superiority of TDD over a test-last approach are in reality due to the fact that most of the experiments employ a \textit{coarse-grained} test-last process---closer to the Waterfall approach---as control group. This creates a large differential in granularity between the treatments. 
In the end, it is possible that TDD performs better only when compared to a coarse-grained development process \cite{Pancur:2011ef}.

\section{Discussion}\label{sec:discussion}
In this section, we answer the research questions and briefly discuss the implications. To answer the research questions, we rely on the results of the multiple regression analysis, focusing on the models presented in Tables \ref{tbl:QLTYb} and \ref{tbl:PRODb} after feature selection.

We remark first that the final models are highly significant, and have a medium-sized adjusted $R^2$. However, the limitations presented in Section \ref{sec:ttv} should not be ignored in operationalizing the results. 

\textbf{RQ-QLTY---Which subset of the four factors best explain external quality?} 
Improvements in external quality were  associated with granular (short) development cycles with minimal variation. Refactoring, counter-intuitively, had a negative association with external quality. Most notably, sequencing---the dimension which is most commonly identified with TDD---was absent in the model. There were no significant interactions between granularity and uniformity. We conclude that granularity, uniformity and refactoring effort together constitute the best explanatory factors for external quality. 

\textbf{RQ-PROD---Which subset of the four factors best explain developer productivity?} 
The results for productivity were similar to those of quality. Improvements in productivity were also associated with the factors related to both cyclical characteristics of the process---granular cycles with minimal variation---and refactoring effort. Sequencing and interactions were absent.   
We conclude that granularity, uniformity and refactoring effort together constitute the best explanatory factors for developer productivity. 
Given the study limitations and the answers to the research questions, we operationalize the results in terms of the following take-aways. 

\textbf{Granularity and uniformity are possibly the most important factors in TDD-like processes.} 
According to our results, emphasizing the iterative nature of the process provides the ``best bang for the buck''. 
We thus recommend focusing on breaking down development tasks into as small and as uniform steps as possible. We think that this aspect should be emphasised over religiously focusing on leading each production cycle with unit tests. Our results confirm the common advice of limiting the length of production (test-code or code-test) and refactoring cycles to five to ten minutes and keeping a steady rhythm.    

Short cycles and a steady rhythm go hand in hand, meaning these two characteristics \textit{together} make a difference. In isolation, they may not be as effective since some of the individual coefficients in the regression models were insignificant. Further studies should focus on the isolated effects of granularity and uniformity, specifically on developer productivity. 

\textbf{The order in which unit tests are written vis-\'a-vis production code does not appear to be important for TDD-like processes.} This finding goes against the common folklore about TDD that stresses leading production cycles with unit tests, above other aspects. Sequencing does not appear among the factors selected for either external quality or productivity in the final models.

The absence of sequencing as an influential dimension does not imply that a strictly \textit{develop-then-test} (test-last) strategy should be preferred over a \textit{test-first} strategy: this advice would require a negative (statistically significant) coefficient, which the models did not produce.
Our result simply states that the order in which unit tests and production code are written may not be as important as commonly thought so long as the process is iterative, granular, and uniform. Therefore, the developers could follow the approach they prefer while paying particular attention to step size and keeping a steady rhythm. 

We further qualify this latter advice with the limitations of the study in mind. A test-first dynamic may provide long-term advantages not addressed by or detected in our study. For example we have not tackled the potential benefits of test-first in resolving requirements uncertainty, formalizing design decisions, and encouraging writing more tests \cite{erdogmus:encyclopedia}, all of which may kick in the longer term and tip the balance in favor of an emphasis for test-first. 


We are not able to draw general or sweeping conclusions regarding refactoring effort due to the validity threat associated with the representation of this construct and the short-term nature of the study. In our study, refactoring effort was negatively associated with  both external quality (an unexpected finding) and developer productivity (an expected finding). However these findings were subject to limitations. We believe that longer-term studies with better constructs are required to reveal the quality and productivity effects of refactoring activity  in the context of TDD-like processes. 
Such effects should be visible once the risk of breaking the system becomes a significant risk and the test assets start paying off in terms of the coverage required to support adding new functionality. Moreover the effects, harmful or beneficial, possibly depend on the nature and distribution of the refactoring activity---e.g., whether \textit{pure} or \textit{floss}--and thus the construct must be represented by measures that are able to differentiate between the different types.  





In summary, dissecting the TDD process into its more elementary aspects shed light on certain established TDD myths. 
Our results suggest that the secret of TDD, in particular with respect to quality, might not be centered on its test-first nature, but rather on its ability to encourage developers to consistently take fine-grained steps, what Kent Beck calls \textit{baby steps} \cite{beck2003test}, provided that they keep writing tests. 
Thus TDD, and its variants, can be thought of as a process that facilitates a highly fine-grained approach to software development, which in turn promises to improve quality and productivity. Such improvements may be small or uncertain in the short term. Further studies are needed to investigate more persistent, long-term effects and the role of refactoring. 

An important corollary of our findings is that incremental test-last (ITL) and TDD are not necessarily that different deep down: they could be substitutes and equally effective provided that they are performed at the same level of granularity and uniformity. Rafique and Misic \cite{rafique2013effects}, Causevic et al. \cite{Causevic:2011hp}, \cite{muller2007effect}, and Pancur et al. \cite{Pancur:2011ef} previously speculated on the importance of granularity, as discussed in Section \ref{sec:background}. Our findings support their intuition. 

We encourage other researchers to attempt to replicate this study, preferably in varying contexts and with different subject profiles. A lab package \cite{Fucci2016} is available for replication purposes.

\section*{Acknowledgments}
This research is supported in part by the Academy of Finland Project \#278354 and the TEKES FiDiPro Programme.
We would like to acknowledge Dr. Lucas Layman, who designed one of the tasks used in the study. We are indebted to the reviewers whose diligence, thoroughness and attention to detail were extremely helpful. They identified many mistakes and led us to uncover inconsistencies in the data. In particular, through their suggestions and insights, we were able to revise and improve our measures and analyses considerably. 
%

%
%
\bibliographystyle{IEEEtran}
\bibliography{bibliography}  

\end{document}